\documentclass[useAMS,usenatbib,letters]{mn2e}
\usepackage{aas_macros,graphicx,times,multirow}
\title[Re-activation of \src]
  {The 2008 May burst activation of SGR\,1627--41}
\author[P. Esposito et al.]
{P.~Esposito,$^{1,2}$\thanks{E-mail: paoloesp@iasf-milano.inaf.it} G.~L.~Israel,$^{3}$ S.~Zane,$^{4}$ F.~Senziani,$^{2,5}$ R.~L.~C.~Starling,$^{6}$
\newauthor N.~Rea,$^{7}$ D.~M.~Palmer,$^{8}$ N.~Gehrels,$^{9}$ A.~Tiengo,$^{2}$ A.~De~Luca,$^{1,2,5}$ D.~G\"otz,$^{10}$
\newauthor S.~Mereghetti,$^{2}$ P.~Romano,$^{11}$ T.~Sakamoto,$^{8}$ S.~D.~Barthelmy,$^{8}$ L.~Stella,$^{3}$
\newauthor R.~Turolla,$^{12,4}$ M.~Feroci,$^{13}$ and V.~Mangano$^{11}$\smallskip\\
$^1$Universit\`a degli Studi di Pavia, Dipartimento di Fisica Nucleare e Teorica and INFN-Pavia, via A.~Bassi 6, 27100 Pavia, Italy\\
$^2$INAF/Istituto di Astrofisica Spaziale e Fisica Cosmica - Milano, via E.~Bassini 15, 20133 Milano, Italy\\
$^3$INAF/Osservatorio Astronomico di Roma, via Frascati 33, 00040 Monteporzio Catone, Italy\\
$^4$University College London, Mullard Space Science Laboratory, Holmbury St. Mary, Dorking, Surrey RH5 6NT, UK\\
$^5$IUSS - Istituto Universitario di Studi Superiori, viale Lungo Ticino Sforza 56, 27100 Pavia, Italy\\
$^6$University of Leicester, Department of Physics and Astronomy, Leicester, LE1 7RH, UK\\
$^7$University of Amsterdam, Astronomical Institute Anton Pannekoek, Kruislaan 403, 1098~SJ Amsterdam, The Netherlands\\
$^8$Los Alamos National Laboratory, Los Alamos, New Mexico 87545, USA\\
$^9$NASA Goddard Space Flight Center, Greenbelt, Maryland 20771, USA\\
$^{10}$CEA Saclay, DSM/Irfu/Service d'Astrophysique, Orme des Merisiers, B\^at. 709, 91191 Gif sur Yvette, France\\
$^{11}$INAF/Istituto di Astrofisica Spaziale e Fisica Cosmica - Palermo, via U.~La Malfa 163, 90146 Palermo, Italy\\
$^{12}$Universit\`a degli Studi di Padova, Dipartimento di Fisica, via F.~Marzolo 8, 35131 Padova, Italy\\
$^{13}$INAF/Istituto di Astrofisica Spaziale e Fisica Cosmica - Roma, via Fosso del Cavaliere 100, 00133 Roma, Italy}
\date{Accepted 2008 July 10. Received 2008 July 10; in original form 2008 June 18}
\pagerange{\pageref{firstpage}--\pageref{lastpage}} \pubyear{2008}
\def\LaTeX{L\kern-.36em\raise.3ex\hbox{a}\kern-.15em
    T\kern-.1667em\lower.7ex\hbox{E}\kern-.125emX}
\def\xmm {\emph{XMM-Newton}}
\def\cxo {\emph{Chandra}}
\def\swift {\emph{Swift}}
\def\sax {\emph{BeppoSAX}}

\def\src {SGR\,1627--41}
\def\flux {\mbox{erg cm$^{-2}$ s$^{-1}$}}
\def\lum {\mbox{erg s$^{-1}$}}

\begin{document}
%\special{!userdict begin /bop-hook{gsave 150 90 translate 55 rotate /Times-Roman findfont 150 scalefont setfont 0 0 moveto 0.9 setgray (Confidential) show grestore}def end}
\label{firstpage}
\maketitle
\begin{abstract}
In May 2008 the soft gamma-ray repeater \src\ resumed its bursting activity after nearly a decade of quiescence. After detection of a bright burst, \swift\ pointed its X-ray telescope in the direction of the source in less than five hours and followed it for over five weeks. In this paper we present an analysis of the data from these \swift\ observations and an \xmm\ one performed when \src\ was still in a quiescent state. The analysis of the bursts detected with \swift/BAT shows that their temporal and spectral properties are similar to those found in previous observations of \src\ and other soft gamma-ray repeaters. The maximum peak luminosity of the bursts was $\sim$$2\times10^{41}$ \lum. Our data show that the outburst was accompanied by a fast flux enhancement and by a hardening of the spectrum with respect to the persistent emission.
\end{abstract}
\section{Introduction}
The soft gamma-ray repeater (SGR) \src\ is likely to host a `magnetar', i.e. an isolated neutron star believed to have an extremely strong magnetic field ($B\sim 10^{14}$--$10^{15}$ G) powering their bright X-ray emission and peculiar bursting activity \citep[e.g.][]{mereghetti08}. Several magnetars, including \src, have been observed to emit short bursts ($<$1 s) in the hard X/soft gamma-ray band, with
characteristic peak luminosities of the order of $\sim$$10^{39}$--$10^{41}$ \lum. Besides short bursts, SGRs are known to emit intermediate and giant flares, with typical durations of 0.5--500 s, during which luminosities up to $\sim$$10^{47}$ \lum\ can be achieved.\\
%The short bursts are only part of the large variety of bursts, flares and outbursts that magnetars can undergo: e.g. the so called intermediate and giant flares, where on time-scales of the orders of 0.5--500 s extremely large energies are emitted, reaching luminosities as high as $\sim$$10^{47}$ \lum.\\
\indent \src\ was discovered in 1998 by the \emph{Compton Gamma Ray Observatory} because of the intense bursts it emitted at the time \citep{kouveliotou98_1627}. These bursts, more than a hundred in six weeks, were also observed by other X-ray satellites \citep{hurley99_1627,woods99,feroci98,smith99,mazets99}. Soon after the discovery of the bursts, the persistent X-ray emission of this SGR was detected by \sax\ at a luminosity of $\sim$$10^{35}$ \lum\ \citep[assuming a distance to the source of 11 kpc;][]{corbel99}. The quiescent spectrum was well modelled by an absorbed power law ($N_{\rm H}\approx8\times10^{22}$ cm$^{-2}$ and photon index $\Gamma\approx2.5$; \citealt{woods99}). No further bursting activity has been reported since then, but several X-ray satellites observed the X-ray persistent counterpart of \src\ in the past ten years \citep{kouveliotou03,mereghetti06}. Since its discovery, this persistent emission showed a slow luminosity decay, from about $10^{35}$ to $10^{33}$ \lum, the lowest value ever observed for an SGR, and a spectral softening from $\Gamma\approx2$ to 4 \citep{kouveliotou03,mereghetti06}. The post-burst cooling trend seen in X-rays is peculiar among SGRs; rather it resembles the behaviour of transient anomalous X-ray pulsars (AXPs), a sub-class of the magnetar family.\\
\indent Here we report on the last observation of \src\ performed at the end of the ten year long stretch of quiescence and on the burst re-activation of the source on 28 May 2008 \citep{palmer08,golenetskii08,woods08atel1549}.
%In particular on the results obtained from the Burst Alert Telescope (BAT), X-Ray Telescope (XRT), Ultraviolet/Optical Telescope (UVOT) on board of \swift\ \citep{gehrels04short}. In Section we report on the analysis of the ... observations, in Section on our results and discussion follows in Section.

%\section{Last observations of \src\ in quiescence}\label{quiescence}
\section{The February 2008 \emph{XMM-Newton} observation}
\begin{figure}
\resizebox{\hsize}{!}{\includegraphics[angle=0]{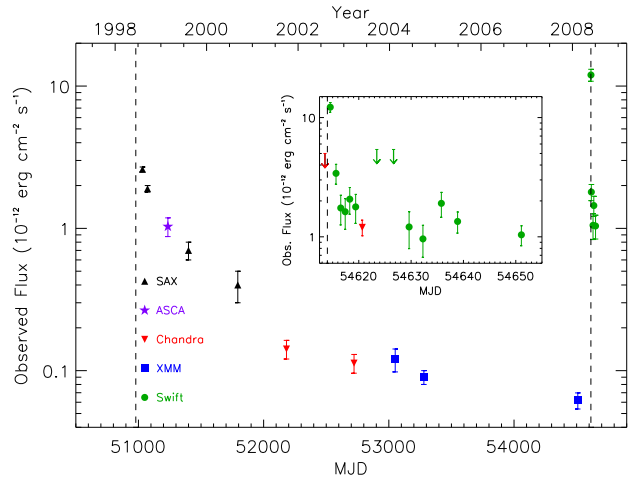}}
\caption{\label{history} Long term light-curve of \src\ based on data from different satellites (updated from \citealt{mereghetti06}). The vertical dashed lines indicate the onset of the two burst-active periods of the source. The inset shows in detail the light-curve around the 2008 reactivation, using also \cxo\ data from \citet{tiengo08atel1559} and \citet{woods08atel1564}. The down-arrows indicate upper limits at 3\,$\sigma$ confidence level.}
\end{figure}

The last \xmm\ \citep{jansen01} observation of \src\ before the May 2008 re-activation was carried out on 12--13 February 2008 and lasted about 80 ks. 
The EPIC pn and MOS cameras (sensitive in the 0.1--15 keV range) were operated in Full Frame mode with the medium optical filter.
%Here we report on the results obtained with the EPIC instrument, that includes one pn and two MOS X-ray CCD cameras (sensitive in the 0.1--15 keV range). All cameras were operated in Full Frame mode and the medium thickness filter was used. 
The data were processed using version 7.1.0 of the \xmm\ Science Analysis Software (\textsc{sas}). We selected events with patterns 0--4 and 0--12 for the pn and the MOS cameras, respectively. The data were filtered to reject intervals with soft-proton flares, reducing the net exposure time to 49.3 ks for the pn detector, 68.3 ks for the MOS\,1, and 69.4 ks for the MOS\,2.\\
\indent Source spectra were accumulated for each camera from circular regions with a 25\arcsec\ radius. The background counts were selected from a $70\times150$ arcsec$^2$ box centred at $\rm RA=16^h36^m01\fs4, Decl.=-47\degr34\arcmin27\farcs6$. 
%Different choices of background regions result in somewhat different best-fit values for the spectral parameters, but do not affect the results reported in this work. 
About 330 counts above the background were collected from \src\ by the pn between 2 and 10 keV, 150 by the MOS\,1, and 160 by the MOS\,2.
Spectral redistribution matrices and ancillary response files were generated using the \textsc{sas} tasks \textsc{rmfgen} and \textsc{arfgen}. The spectral fitting was performed using the \textsc{xspec} fitting package version 12.4. The data were grouped so as to have at least 20 counts per energy bin and the spectra from the MOS\,1, MOS\,2, and pn in the 2--10 keV range were fit simultaneously (spectral channels having energies below 2 keV were ignored, owing to the very low signal-to-noise ratio). We fit an absorbed power-law model and obtained the following best-fit parameters ($\chi^2_{\rm{r}}=0.98$ for 42 degrees of freedom): absorption $N_{\rm H}=(10\pm2)\times10^{22}$ cm$^{-2}$ and photon index \mbox{$\Gamma=3.3^{+0.6}_{-0.4}$} (hereafter all errors are at 1\,$\sigma$ confidence level). The observed 2--10 keV flux was $(6\pm2)\times10^{-14}$ \flux. For comparison with previous work \citep{kouveliotou03,mereghetti06} we also fit the data keeping the absorption column fixed at $N_{\rm H}=9\times10^{22}$ cm$^{-2}$, obtaining a similar photon index \mbox{$\Gamma=3.0\pm0.2$} and a flux of $(6\pm1)\times10^{-14}$ \flux. This value  (plotted in Figure~\ref{history}) hints at a further luminosity decrease since the previous \xmm\ observations \citep[September 2004;][]{mereghetti06}.
% The spectra could also be fit by a blackbody with $kT\simeq1$ keV and emetting radius of $\sim$0.1 km, corrected for interstellar absorption ($N_{\rm H}\simeq4\times10^{22}$ cm$^{-2}$).

% \subsection{The May 2008 serendipitous \emph{Chandra} observation}\label{cxo-ul}
% Following the Swift detection of the reactivation of \src\ \citep{palmer08,golenetskii08,woods08atel1549}, we searched for serendipitous pre-bursts X-ray observations of the region. \cxo\ observed the field of this SGR on 2008-05-27 UT 12:57:00 (Obs-ID 9007), $\sim$20 hours before the first burst detected by BAT from \src. The \cxo\ observation was performed using HRC-I with an exposure time of $\sim$1.2 ks, and the SGR position was at $\sim$15\arcmin\ from the aim-point of the observation.\\
% \indent \src\ was not detected in this pre-bursts \cxo\ observation. Taking into account the off-axis position of the source and point-spread function blurring, and assuming an absorbed power-law spectrum ( $N_{\rm H}=9\times10^{22}$ cm$^{-2}$ and $\Gamma=3.2$), we infer a 3\,$\sigma$ upper limit on the absorbed flux of  $5\times10^{-12}$ \flux\ (0.5--10 keV band). 

\section{\emph{Swift} observations and data analysis}
\swift\ \citep{gehrels04short} was specifically designed to study Gamma Ray Bursts (GRBs) and their afterglows, and its payload includes a wide-field instrument, the Burst Alert Telescope \citep[BAT;][]{barthelmy05}, and two narrow-field instruments, the X-Ray Telescope \citep[XRT;][]{burrows05short} and the Ultraviolet/Optical Telescope \citep[UVOT;][]{roming05short}. In this Section we report on the results obtained from our analysis of the \swift\ BAT and XRT observations of \src\ performed since its May 2008 re-activation.
Given the extremely large optical extinction inferred from the X-ray absorption of the \src\ spectrum \citep[$A_{\rm V}> 40$ mag;][]{wachter04} the UVOT instrument cannot provide meaningful constraints on the ultraviolet/optical emission of \src.
\subsection{Burst Alert Telescope data - Bursting emission}\label{bat}
\begin{figure*}
 \begin{minipage}[t]{\hsize}
\resizebox{\hsize}{!}{\includegraphics[angle=0]{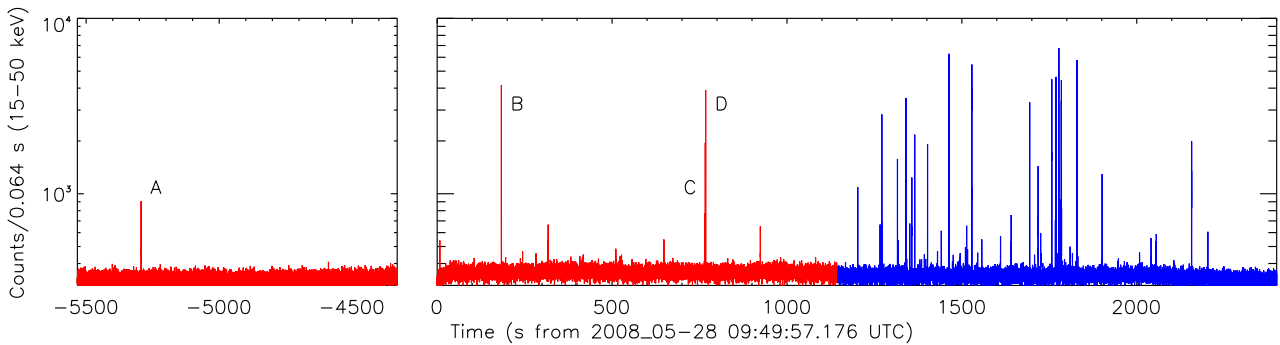}}
\caption{\label{forest} The time-line of the bursts detected by the \swift\ Burst Alert Telescope (generated on board, see Section \ref{bat}). The portion plotted in red corresponds to the temporal range for which event data were available; for the portion in blue only rate and survey data were distributed.}
 \end{minipage}
\end{figure*}

\begin{figure}
\centering
\resizebox{.9\hsize}{!}{\includegraphics[angle=-90]{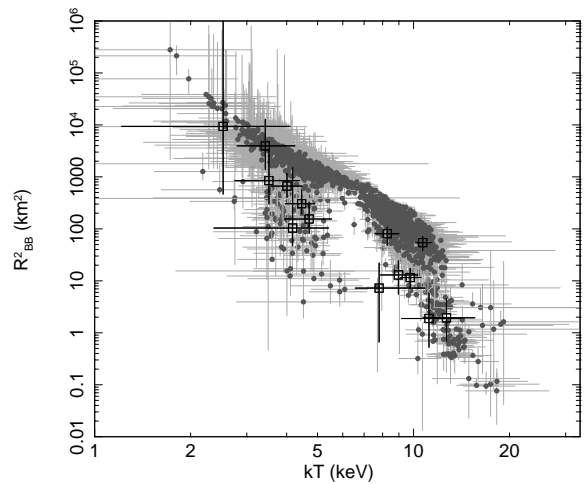}}
\caption{\label{2bb} Square of radii for the 2BB model as a function of the corresponding temperatures for the time-resolved BAT data of bursts: empty squares mark the \src\ events detected in 2008 May, grey data refer to the 2006 March `burst forest' of SGR\,1900+14 \citep{israel08short}.}
\end{figure}

The coded mask gamma-ray (15--150 keV) BAT instrument spends a large fraction of its time waiting for the occurrence of a GRB in its field of view (FOV). Whenever a GRB or an interesting hard X-ray transient is detected, information for individual photons is sent to the ground in order to have the maximum energy and time resolution (event data). If no GRB is detected, the on board software accumulates the detector count map in 80-channel histograms with a typical integration time of $\sim$5 minutes (survey data). %
%In this mode, light-curves for bright sources in the FOV are generated by the on-board software.\\
In this mode, continuous full-detector count rate information is available in 4 energy bands at 64 ms resolution, providing a light-curve for bright variable sources in the FOV.\\
\indent On 28 May 2008 at 08:21:43 UT BAT triggered on and localised a bright burst from \src\ \citep{palmer08}. Another  bright burst was detected at 09:53:00 UT and was followed by tens of bursts extending to at least 10:25:54 UT (see Figure~\ref{forest}).
Due to the non-continuous coverage, the net exposure time spent by BAT on the source was $\sim$3.4 ks. 
In this Section we present the study of bursts detected when event data were available ($\sim$2.3 ks).
%Where event data were available, a detailed study of single bursts was possible, otherwise average spectral properties were deduced using survey data.
\\
\indent The data reduction was performed using version 2.8 of the standard \swift/BAT analysis software distributed within \textsc{ftools} under the \textsc{heasoft} package (version 6.4.0). For each event file, the background-subtracted counts of the source were extracted from the detector pixels illuminated by the source by using the mask-weighting technique.
%\footnote{See http://heasarc.nasa.gov/docs/swift/analysis/threads/batfluxunitsthread.html.} 
Light-curves in the 15--50 keV band showing the bursting activity were produced. For each burst, a spectrum of the entire bursting time interval was extracted.\\
\indent We identified in the BAT data eight bursts with more than 500 counts in the 15--50~keV  energy range. Their spectra were fit well by an optically-thin thermal bremsstrahlung (OTTB) model with temperatures ranging from about 10 to 70 keV and are overall similar to those detected during the previous outburst of \src\ as well as the outbursts from other SGRs \citep[][]{aptekar01}.\\ 
\indent The main problem with the OTTB model is that, while it generally provides good fits to the spectra of SGR bursts in the hard X-rays ($\ga$15 keV), it tends to overestimate the flux at low energies when broad band data are available \citep[e.g.][]{olive04,feroci04}. Among numerous possible spectral models \citep[see][for a review]{israel08short}, we tested the double blackbody model (2BB) that was successfully applied to the SGR bursting emission over broad energy ranges \citep[e.g.][]{olive04,feroci04,nakagawa07short,israel08short}. In the case of \src\ the 2BB model yielded a good description of the burst spectra as well (even though the additional free parameters with respect to the OTTB model were not statistically required); in Table \ref{bat-fits} we show the spectral results for the brightest events. Similar to previous studies, our spectral fits show the presence of a `cold' blackbody with $kT_1\approx4$ keV and emitting radius $R_1\approx30$ km and a hotter blackbody, with $kT_2\approx10$ keV and $R_2\approx4$ km.\\
\begin{table*}
\centering
\begin{minipage}{16.5cm}
\caption{Spectral fit results (in the 15--100 keV range) of the bright bursts detected by \swift/BAT (see Figure~\ref{forest}) for the double blackbody model. We assume a distance to the source of 11 kpc.}
\label{bat-fits}
\centering
\begin{tabular}{@{}ccccccccccc}
\hline
Burst & Net counts & Duration & $kT_1$ & $R_1$ & $L_1$$^{\rm a}$ & $kT_2$ & $R_2$ & $L_2$$^{\rm a}$ & Fluence & $\chi^{2}_{r}$ (d.o.f.)\\
   & & (s) & (keV) & (km) & (\lum) & (keV) & (km) & (\lum) & (erg cm$^{-2}$) & \\
\hline
A & 11\,802 & 0.110 & $3.8\pm0.7$ & $35^{+26}_{-12}$ & $3.2\times10^{40}$ & $8.4\pm0.9$ & $5\pm2$ & $1.6\times10^{40}$ & $2.0\times10^{-7}$  & 0.90 (34)\\
% & & $5.8^{+1.0}_{-0.8}$ & $9^{+4}_{-3}$ & -- & -- & $1.8\times10^{-7}$  & 1.06 (36)\\
% & & $16^{+2}_{-3}$ & --& --& --& $2.0\times10^{-7}$  & 0.92 (36)\\
B &  20\,595 & 0.135 & $4.0^{+0.5}_{-0.4}$ & $30^{+10}_{-7}$ & $3.1\times10^{40}$ & $10.7^{+0.7}_{-0.5}$ & $3.9\pm0.6$ & $2.6\times10^{40}$ & $3.5\times10^{-7}$  & 0.91 (34)\\
% & & $28\pm2$ & --& --& --& $3.5\times10^{-7}$  & 0.92 (36)\\
C & 3\,704 & 0.018 & $3.9^{+0.8}_{-0.7}$ & $34^{+22}_{-12}$ & $3.6\times10^{40}$ & $11^{+2}_{-1}$ & $4\pm1$ & $4\times10^{40}$ & $0.6\times10^{-7}$  & 1.22 (34)\\
% & & \textsc{ottb} & $36^{+8}_{6}$ & --& --& --& $0.6\times10^{-7}$  & 1.21 (36)\\
D & 10\,497 & 0.250 & $4.4\pm0.5$ & $15^{+4}_{-3}$ & $1.1\times10^{40}$ & $10\pm1$ & $2.0^{+0.9}_{-0.7}$ & $0.5\times10^{40}$ & $1.7\times10^{-7}$  & 1.07 (34)\\
% & & $19\pm2$ & --& --& --& $1.7\times10^{-7}$  & 1.10 (36)\\
\hline
\end{tabular}

\begin{list}{}{}
\item[$^{\rm a}$] Bolometric luminosity of first/second blackbody.
\end{list}
\end{minipage}
\end{table*}
%\begin{table}
%\centering
%\caption{Spectral fit results (in the 15--100 keV range) of bursts detected by \swift/BAT for the OTTB model.}
%\label{bat-fits-brem}
%\centering
%\begin{tabular}{@{}ccccc}
%\hline
%Burst$^{\rm a}$ & Net counts &  $kT$ & Fluence & $\chi^{2}_{r}$ (d.o.f.)\\
%                            &                      & (keV)      &  (erg cm$^{-2}$) & \\
%\hline
%7 s      & 1\,203 & $18^{+14}_{-7}$ & $0.1\times10^{-7}$  & 1.13 (39)\\
%316 s & 1\,179 & $13^{+10}_{-5}$ & $0.1\times10^{-7}$  & 0.59 (36)\\
%765 2 & 1\,524 & $20^{+8}_{-6}$   &  $0.2\times10^{-7}$  & 1.04 (36)\\
%766 s & 3\,704 & $37^{+8}_{-7}$   &  $1.0\times10^{-7}$ & 1.22 (36)\\
%924 s & 1\,146 & $17^{+9}_{-6}$   &  $0.1\times10^{-7}$  & 1.04 (36)\\
%\hline
%\end{tabular}
%
%\begin{list}{}{}
%\item[$^{\rm a}$] Errors are given at the 90\% confidence level.
%\end{list}
%\end{table}\\
\indent For these bright bursts we also produced the spectra corresponding to the rise, peak, and decay phases. The maximum luminosity detected in the \src\ data set was $\sim$$2\times10^{41}$ \lum\ (burst B, peak phase; in particular, the hard blackbody component, with $kT_2\simeq11$ keV and radius $R_2\simeq8$ km, reached a luminosity of $\sim$$10^{41}$ \lum ). Small variations with time were detected, though all the spectra are consistent with the model parameters of the corresponding time-averaged spectrum (see Table \ref{bat-fits}) simply re-scaled in normalisation (to account for the luminosity evolution during the burst). The results of this analysis are reported in Figure~\ref{2bb} where the two blackbody equivalent surfaces are shown as a function of their temperatures (only time resolved spectra with well-constrained fitting parameter values are shown). For comparison, we also report the corresponding measurements obtained for the intense `burst forest' emitted by SGR\,1900+14 on 29 March 2006 and observed by BAT \citep{israel08short}. Despite the small number of photons detected in the \src\ bursts, their spectral properties are in good agreement with those of SGR\,1900+14.
\\
\indent We also searched the data for a persistent emission from \src\ during all the non-bursting intervals. For each event file, an image was generated excluding the burst time intervals and the \textsc{batcelldetect} tool was run. We investigated two time intervals, the first one (see Figure~\ref{forest}) ranging from time $t=-5\,533$ to $-4\,331$ s (net exposure time 1\,201 s) and the second one from $t=0$ to $1\,146$ s (net exposure  1\,140 s). In both cases we found no significant emission; the 3\,$\sigma$ upper limits on the flux in the 15--50 keV band for the above quoted intervals are $10^{-9}$ and $4\times10^{-10}$ \flux, respectively (the large difference is due to the coded fraction at which the source was observed in the two cases). 

\subsection{X-Ray Telescope data - Persistent emission}
The \swift\ XRT uses a CCD detector sensitive to photons with energies between 0.2 and 10 keV. Fourteen \swift\ observations of \src\ were performed following the source re-activation. The XRT instrument was operated in photon counting (PC) mode. The first observation started about 4.6 hours after the first burst was detected by the BAT. Table \ref{log} reports the log of the XRT observations used for this work.
\begin{table}
\centering
\caption{Journal of the 2008 \swift/XRT observations. The observation sequence number is composed of 00312579 followed by the three digit segment number given here (e.g. 00312579001).}
\label{log}
\begin{tabular}{@{}cccccc}
\hline
Obs. & Date & \multicolumn{2}{c}{Start/End time (UT)} & Exposure$^{\rm a}$ & Count rate\\
 & mm-dd & \multicolumn{2}{c}{hh:mm:ss} & (ks) & (counts s$^{-1}$)\\
\hline
001 & 05-28 & 12:58:14 & 13:31:27 & 2.0 & $0.067 \pm 0.003$\\
002 & 05-29 & 14:47:17 & 16:30:39 & 2.0 & $0.015 \pm 0.003$\\
003 & 05-30 & 11:35:15 & 14:58:56 & 1.9 & $0.010 \pm 0.003$\\
004 & 05-31 & 08:49:41 & 10:37:58 & 1.8 & $0.007 \pm 0.002$\\
005 & 06-01 & 05:32:11 & 07:34:56 & 2.0 & $0.008 \pm 0.002$\\
006 & 06-02 & 08:37:17 & 19:18:48 & 2.1 & $0.011 \pm 0.003$\\
007 & 06-06 & 10:35:23 & 12:16:30 & 0.6 & $<$0.03$^{\rm b}$\\
008 & 06-09 & 14:41:22 & 16:19:56 & 0.3 & $<$0.03$^{\rm b}$\\
009 & 06-12 & 13:07:43 & 16:35:58 & 1.9 & $0.006 \pm 0.002$\\
010 & 06-15 & 03:34:10 & 08:50:57 & 3.8 & $0.006 \pm 0.002$\\
011 & 06-18 & 16:54:48 & 20:24:56 & 2.3 & $0.010 \pm 0.002$\\
012 & 06-21 & 18:46:38 & 23:39:55 & 5.2 & $0.008 \pm 0.002$\\
013 & 07-02 & 05:17:00 & 17:10:56 & 1.5 & $0.005 \pm 0.002$\\
014 & 07-05 & 15:29:07 & 21:59:57 & 5.6 & $0.005 \pm 0.001$\\
\hline
\end{tabular}
\begin{list}{}{}
\item[$^{\rm a}$] The exposure time is usually spread over several snapshots (single continuous pointings at the target) during each observation.
\item[$^{\rm b}$] Upper limit at 3\,$\sigma$ confidence level \citep[following][]{gehrels86}.
\end{list}
\end{table}
The data were processed with standard procedures using the \textsc{ftools} task \textsc{xrtpipeline} (version 0.11.6). We selected events with grades \mbox{0--12} and limited the analysis to the 0.3--10 keV range, where the PC response matrices are well calibrated.\\
\indent The source was significantly detected in all observations with exposure time longer than 1 ks with the mean count rates given in Table \ref{log}. These values, corrected for the loss of counts caused by hot columns and pixels (the correction factor was calculated with \textsc{xrtmkarf}), are plotted in Figure~\ref{history} after conversion to flux as described below.\\
\indent For the spectral analysis, we extracted the source events from a circular region with a radius of 20 pixels (1 pixel $\simeq 2.37''$), whereas the background events were extracted within an annular source-free region centred on \src\ and with radii 50 and 75 pixels. 
%The choice of the XRT background region has no effect in the spectral analysis, since it is only $<$9\% of the source net counts.
%Since the data show a maximum count rate $<$0.1 counts s$^{-1}$, no pile-up correction was necessary. 
Since individual data sets have too few counts for a meaningful spectral analysis, we extracted a cumulative spectrum. This corresponds to a total exposure of 32.9 ks and contains about 340 net counts in the 0.3--10 keV range. The data were rebinned with a minimum of 15 counts per energy bin. The ancillary response file was generated with \textsc{xrtmkarf}, and it accounts for different extraction regions, vignetting and point-spread function corrections. We used the latest available spectral redistribution matrix (swxpc0to12s6\_20010101v010.rmf).
Adopting an absorbed power-law model, we find the following best-fit parameters ($\chi^2_{\rm{r}}=1.13$ for 21 d.o.f.): absorption $ N_{\rm H}=10^{+4}_{-3}\times10^{22}$ cm$^{-2}$ and photon index $\Gamma=1.5^{+0.7}_{-0.4}$. We used the resulting observed 2--10 keV flux of $\sim$$2.3\times10^{-12}$ \flux\ in order to derive the conversion factor \mbox{1 count s$^{-1}$ $\simeq$ $1.8 \times10^{-10}$ \flux}.\\
%An absorbed ($N_{\rm H}=(6\pm4)\times10^{22}$ cm$^{-2}$) blackbody model with $kT=1.8^{+0.8}_{-0.4}$ keV and radius $R=0.2^{+0.2}_{-0.1}$ km provides an equally good fit ($\chi^2_{\rm{red}}=1.04$ for 14 d.o.f.). The corresponding 0.3--10 keV observed flux is $\sim$$3.0\times10^{-12}$ \flux.\\
\indent Although the uncertainties in the spectral parameters are large,  the \swift/XRT spectrum appears to be harder than the one observed by \xmm/EPIC in February 2008. To better investigate whether this hardening is statistically significant, we simultaneously fit the XRT and EPIC spectra with an absorbed power-law model with linked parameters and a free normalisation factor.
%, to allow for the different flux observed during the two observations. 
The resulting $\chi^2_{r}$ (1.60 for 65 d.o.f.) is unacceptable, while an acceptable fit ($\chi^2_{r}=1.01$ for 64 d.o.f.) is obtained once the photon index is also left free to vary independently. 
%The best-fit parameters of the latter fit are: a common absorption of $N_{\rm H}=(9^{+4}_{-2})\times10^{22}$ cm$^{-2}$ and photon indices $\Gamma_{\rm XMM}=3.2^{+1}_{-0.5}$ and $\Gamma_{\rm XRT}=1.3^{+0.8}_{-0.6}$ for the \xmm\ and XRT spectra, respectively.
The best-fit parameters of the latter fit are: a common absorption of $N_{\rm H}=(10^{+1}_{-2})\times10^{22}$ cm$^{-2}$ and photon indices $\Gamma_{ \rm XMM}=3.5^{+0.1}_{-0.5}$ and $\Gamma_{\rm XRT}=1.5^{+0.3}_{-0.5}$ for the EPIC and XRT spectra, respectively.

\section{Discussion and conclusions}\label{disc}
\begin{figure}
\resizebox{\hsize}{!}{\includegraphics[angle=-90]{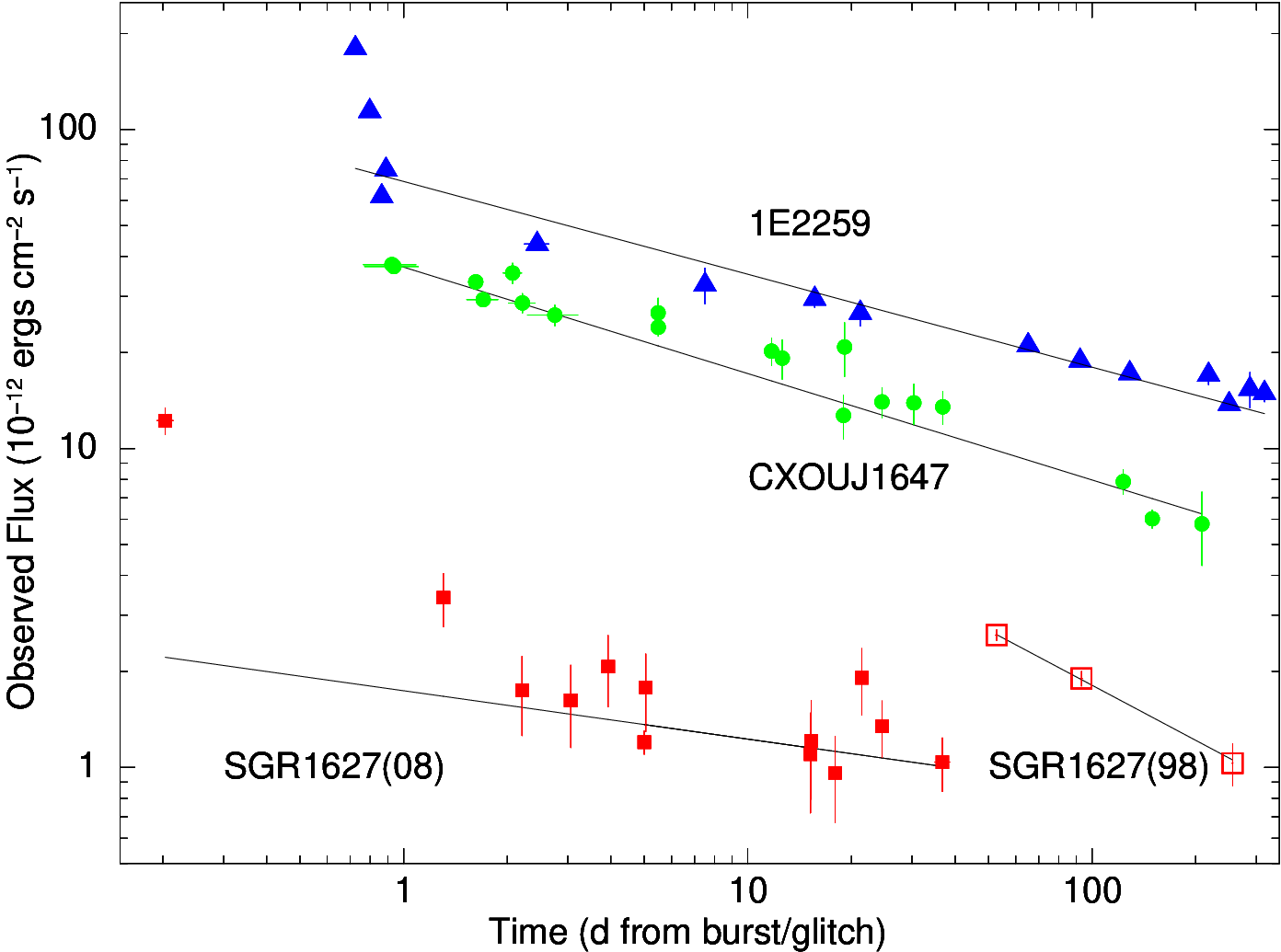}}
\caption{\label{comp} Comparison among flux decays of \src\ (this work, \citealt{woods08atel1564}, \citealt{mereghetti06}), CXOU\,J164710.2--455216 (\citealt[][]{icd07}; filled circles)  and 1E\,2259+586 (\citealt[][]{woods04}; filled triangles) following a period of bursts emission and/or glitch. In the case of \src\ we report the available data for both the 2008 (filled squares) and 1998 (open squares) activation periods. The solid lines represent the power-law best-fits (see Section \ref{disc} for more details).}
\end{figure}

The recent, spectacular re-activation of \src, following a quiescent interval of nearly a decade, triggered the BAT instrument on board \swift\ on 28 May 2008. Tens of bursts were observed, with fluxes exceeding the underlying continuum by a factor $>$$10^5$. The bursts achieved a maximum luminosity of $\sim$$10^{41}$ \lum\ and had a duration of $<$0.5 s, typical of the bursts detected in SGRs. Thanks to the rapid response of \swift, the source was repeatedly observed with XRT in the days following the `burst forest' emission, leading to the earliest post-burst observations ever obtained for this SGR. In fact, at the time of the previous active period, the persistent emission was observed only one month after the first burst detection \citep{woods99}.\\ 
\indent With respect to the last pre-burst \xmm\ observation, the source was detected in May--June 2008 at a much larger flux level (see Figure~\ref{history}) and with a considerably harder spectrum. A serendipitous \cxo\ observation performed only 20 hours before the detection of the bursting activity provides a 3\,$\sigma$ upper limit on the absorbed flux of $5 \times 10^{-12}$ \flux\ \citep[0.5--10 keV band;][]{tiengo08atel1559}, showing that most of the flux enhancement occurred in less than a day. These facts indicate a significant phase transition marked by the burst activation. The correlated spectral hardening/flux increase is in line with what is observed in the long term evolution of other magnetars \citep[e.g.][]{mereghetti08}, and expected in models in which the non thermal X-ray emission is due to resonant up-scattering by magnetospheric currents \citep{tlk02}. \\
\indent The early flux decay of the source is shown in Figure~\ref{comp}, where, for comparison, we also plot the flux of the decay that followed the past bursting activity of \src\ and those of two AXPs, CXOU\,J164710.2--455216 \citep{icd07} and 1E\,2259+586 \citep{woods04}. The \src\ data taken after more than two days from the May 2008 trigger are well fit by a power-law decay (index $\sim-0.2$), but the XRT points at earlier times shows a marked excess over this trend, indicating a very steep initial decay. This behaviour closely resembles that of 1E\,2259+586: in that case, after the June 2002 bursts active phase the source flux showed a double component decay, with a steep component that decayed rapidly during the first $\sim$2 days, followed by a slower year-long decay phase \citep{woods04}. Interestingly, the phase of steep flux decay (and harder spectra) was associated with a long-lasting period of bursting activity. This is consistent with what was observed in CXOU\,J164710.2--455216 and \src: in the former case, in which no steep/prompt decay was observed, the bursting activity was already over at the time of the first observation of the persistent flux. In the latter case, after the activation of June 1998 only a shallow decay was monitored $>$60 days from the first bursts. The light-curve of the recent \src\ decay reveals both, the steep and shallow phase, and the last burst was detected by Konus/\emph{Wind} between the first and the second XRT pointing \citep{golenetskii08}.
% The new observations support the idea that the presence of two flux components may be a common characteristic of all magnetar outbursts, although their relative energetics and the relation between the energy released in tail and bursts can differ dramatically from source to source \citep{woods04}.
The new observations support the presence of two distinct time-scales (a short, $\sim$1 day, and a longer one, $\sim$month) in the flux decay following the outburst of a magnetar.
In this respect, we notice that while a steep decay seems to point toward a magnetospheric effect (for instance following current dissipation or other forms of activity), the decay index of the shallow phase ranges from about $-0.6$ (in the case of \src, 1998) to $-0.2$ (for 1E\,2259+586 and \src, 2008) and is roughly compatible with crustal cooling (considering the uncertainties in the theoretical models, see \citealt{eichler06} and references therein).
The energetics and relative importance of the steep/shallow decay phases, nevertheless, greatly vary from source to source.
Indeed, in some cases this requires a powering mechanism for the tail emission much more energetic than the bursts energy deposition. This might be associated with magnetospheric current dissipation or crustal cooling following an impulsive heat deposition. 
\\ 
\indent For several bursts we had enough counts to perform a spectral analysis. 
%Time resolved investigation only showed a subtle spectral evolution. We considered a few models including OTTB and the 2BB models that have been previously applied in the case of SGR\,1900+14 \citep[see e.g.][]{israel08short}.
The parameter values derived from the 2BB fits are in agreement with the results of a comprehensive analysis of a `burst forest' emitted by SGR\,1900+14 \citep{israel08short}, which showed that the bursts populate almost homogeneously all temperatures between $\sim$2 and $\sim$12 keV, with a bimodal distribution behaviour and a sharp edge in the \mbox{$kT$--$R^2$} plane. This supports the idea of two distinct emitting regions, a cold and larger one and a hot and smaller one which in turn may be associated to the escaping regions of two populations of photons, from the O- and E- polarisation modes. Interestingly, the bright bursts detected from \src\ lie within the cloud of the SGR\,1900+14 bursts (Figure~\ref{2bb}) and the luminosities of the two blackbody components are in agreement with the relation shown in Figure~6 of \citet{israel08short}, suggesting that short bursts form a continuum in terms of spectral properties, duration and fluence. Finally, we notice that in the scenario proposed by \citet{israel08short}, the luminosity of the hot blackbody is not expected to exceed the magnetic Eddington luminosity \citep{thompson95}. In the case of \src, where the maximum luminosity observed by the BAT for the hot blackbody is $\sim$$10^{41}$ \lum, this translates into a lower limit for the magnetic field of $B>1.8\times10^{14}$ G.

\section*{acknowledgements}
This research is based on observations with the NASA/UK/ASI \swift\ mission. We thank the \swift\ duty scientists and science planners for making these observations possible. We also used data obtained with \xmm, an ESA science mission with instruments and contributions directly funded by ESA Member States and NASA. The Italian authors acknowledge the partial support from ASI (ASI/INAF contracts I/088/06/0 and AAE~TH-058). SZ and RLCS acknowledge support from STFC. NR is supported by an NWO Veni Fellowship. DG acknowledges the CNES for financial support.

\vspace{-.2cm}

\bibliographystyle{mn2e}
\bibliography{biblio}

\begin{thebibliography}{}

\bibitem[\protect\citeauthoryear{{Aptekar}, {Frederiks}, {Golenetskii},
  {Il'inskii}, {Mazets}, {Pal'shin}, {Butterworth} \& {Cline}}{{Aptekar}
  et~al.}{2001}]{aptekar01}
{Aptekar} R.~L.,  {Frederiks} D.~D.,  {Golenetskii} S.~V.,  {Il'inskii} V.~N.,
  {Mazets} E.~P.,  {Pal'shin} V.~D.,  {Butterworth} P.~S.,    {Cline} T.~L.,
  2001, \apjs, 137, 227

\bibitem[\protect\citeauthoryear{{Barthelmy}, {Barbier}, {Cummings},
  {Fenimore}, {Gehrels}, {Hullinger}, {Krimm}, {Markwardt}, {Palmer},
  {Parsons}, {Sato}, {Suzuki}, {Takahashi}, {Tashiro} \& {Tueller}}{{Barthelmy}
  et~al.}{2005}]{barthelmy05}
{Barthelmy} S.~D. et~al.,  2005, Space Science Reviews, 120, 143

\bibitem[\protect\citeauthoryear{{Burrows}, {Hill}, {Nousek}, {Kennea},
  {Wells}, {Osborne}, {Abbey}, {Beardmore} \& {et~al.}}{{Burrows}
  et~al.}{2005}]{burrows05short}
{Burrows} D.~N. et~al., 2005, Space
  Science Reviews, 120, 165

\bibitem[\protect\citeauthoryear{{Corbel}, {Chapuis}, {Dame} \&
  {Durouchoux}}{{Corbel} et~al.}{1999}]{corbel99}
{Corbel} S.,  {Chapuis} C.,  {Dame} T.~M.,    {Durouchoux} P.,  1999, \apjl,
  526, L29

\bibitem[\protect\citeauthoryear{{Eichler}, {Lyubarsky}, {Kouveliotou} \&
  {Wilson}}{{Eichler} et~al.}{2006}]{eichler06}
{Eichler} D.,  {Lyubarsky} Y.,  {Kouveliotou} C.,    {Wilson} C.~A.,  2006,
  preprint (astro-ph/0611747)

\bibitem[\protect\citeauthoryear{{Feroci}, {Caliandro}, {Massaro}, {Mereghetti}
  \& {Woods}}{{Feroci} et~al.}{2004}]{feroci04}
{Feroci} M.,  {Caliandro} G.~A.,  {Massaro} E.,  {Mereghetti} S.,    {Woods}
  P.~M.,  2004, \apj, 612, 408

\bibitem[\protect\citeauthoryear{{Feroci}, {Costa}, {Amati}, {Piro}, {Martino},
  {di Ciolo}, {Coletta} \& {Frontera}}{{Feroci} et~al.}{1998}]{feroci98}
{Feroci} M.,  {Costa} E.,  {Amati} L.,  {Piro} L.,  {Martino} B.,  {di Ciolo}
  L.,  {Coletta} A.,    {Frontera} F.,  1998, \iaucirc, 6945, 3

\bibitem[\protect\citeauthoryear{{Gehrels}}{{Gehrels}}{1986}]{gehrels86}
{Gehrels} N.,  1986, \apj, 303, 336

\bibitem[\protect\citeauthoryear{{Gehrels}, {Chincarini}, {Giommi}, {Mason},
  {Nousek}, {Wells}, {White}, {Barthelmy} \& {et~al.}}{{Gehrels}
  et~al.}{2004}]{gehrels04short}
{Gehrels} N. et~al., 2004, \apj,
  611, 1005

\bibitem[\protect\citeauthoryear{{Golenetskii}, {Aptekar}, {Mazets},
  {Pal'shin}, {Frederiks}, {Oleynik}, {Ulanov} \& {Cline}}{{Golenetskii}
  et~al.}{2008}]{golenetskii08}
{Golenetskii} S.,  {Aptekar} R.,  {Mazets} E.,  {Pal'shin} V.,  {Frederiks} D.,
   {Oleynik} P.,  {Ulanov} M.,    {Cline} T.,  2008, GCN Circ., 7778

\bibitem[\protect\citeauthoryear{{Hurley}, {Kouveliotou}, {Woods}, {Mazets},
  {Golenetskii}, {Frederiks}, {Cline} \& {van Paradijs}}{{Hurley}
  et~al.}{1999}]{hurley99_1627}
{Hurley} K.,  {Kouveliotou} C.,  {Woods} P.,  {Mazets} E.,  {Golenetskii} S.,
  {Frederiks} D.~D.,  {Cline} T.,    {van Paradijs} J.,  1999, \apjl, 519, L143

\bibitem[\protect\citeauthoryear{{Israel}, {Campana}, {Dall'Osso}, {Muno},
  {Cummings}, {Perna} \& {Stella}}{{Israel} et~al.}{2007}]{icd07}
{Israel} G.~L.,  {Campana} S.,  {Dall'Osso} S.,  {Muno} M.~P.,  {Cummings} J.,
  {Perna} R.,    {Stella} L.,  2007, \apj, 664, 448

\bibitem[\protect\citeauthoryear{{Israel}, {Romano}, {Mangano}, {Dall'Osso},
  {Chincarini}, {Stella}, {Campana}, {Belloni} \& {et~al.}}{{Israel}
  et~al.}{2008}]{israel08short}
{Israel} G.~L. et~al., 2008, \apj, in press

\bibitem[\protect\citeauthoryear{{Jansen}, {Lumb}, {Altieri}, {Clavel}, {Ehle},
  {Erd}, {Gabriel}, {Guainazzi}, {Gondoin}, {Much}, {Munoz}, {Santos},
  {Schartel}, {Texier} \& {Vacanti}}{{Jansen} et~al.}{2001}]{jansen01}
{Jansen} F. et~al.,  2001, \aap, 365,
  L1

\bibitem[\protect\citeauthoryear{{Kouveliotou}, {Kippen}, {Woods},
  {Richardson}, {Connaughton} \& {McCollough}}{{Kouveliotou}
  et~al.}{1998}]{kouveliotou98_1627}
{Kouveliotou} C.,  {Kippen} M.,  {Woods} P.,  {Richardson} G.,  {Connaughton}
  V.,    {McCollough} M.,  1998, \iaucirc, 6944, 2

\bibitem[\protect\citeauthoryear{{Kouveliotou}, {Eichler}, {Woods},
  {Lyubarsky}, {Patel}, {G{\" o}{\u g}{\" u}{\c s}}, {van der Klis}, {Tennant},
  {Wachter} \& {Hurley}}{{Kouveliotou} et~al.}{2003}]{kouveliotou03}
{Kouveliotou} C. et~al.,  2003, \apjl, 596, L79

\bibitem[\protect\citeauthoryear{{Mazets}, {Aptekar}, {Butterworth}, {Cline},
  {Frederiks}, {Golenetskii}, {Hurley} \& {Il'Inskii}}{{Mazets}
  et~al.}{1999}]{mazets99}
{Mazets} E.~P. et~al.,
  1999, \apjl, 519, L151

\bibitem[\protect\citeauthoryear{{Mereghetti}}{{Mereghetti}}{2008}]{mereghetti%
08}
{Mereghetti} S.,  2008, \aapr, in press

\bibitem[\protect\citeauthoryear{{Mereghetti}, {Esposito}, {Tiengo}, {Turolla},
  {Zane}, {Stella}, {Israel}, {Feroci} \& {Treves}}{{Mereghetti}
  et~al.}{2006}]{mereghetti06}
{Mereghetti} S. et~al.,  2006, \aap, 450,
  759

\bibitem[\protect\citeauthoryear{{Nakagawa}, {Yoshida}, {Hurley}, {Atteia},
  {Maetou}, {Tamagawa}, {Suzuki}, {Yamazaki} \& {et~al.}}{{Nakagawa}
  et~al.}{2007}]{nakagawa07short}
{Nakagawa} Y.~E. et~al., 2007, \pasj, 59, 653

\bibitem[\protect\citeauthoryear{{Olive}, {Hurley}, {Sakamoto}, {Atteia},
  {Crew}, {Ricker}, {Pizzichini}, {Barraud} \& {Kawai}}{{Olive}
  et~al.}{2004}]{olive04}
{Olive} J.-F. et~al., 2004, \apj,
  616, 1148

\bibitem[\protect\citeauthoryear{{Palmer}, {Esposito}, {Barthelmy}, {Cummings},
  {Gehrels}, {Israel}, {Krimm}, {Sakamoto} \& {Starling}}{{Palmer}
  et~al.}{2008}]{palmer08}
{Palmer} D. et~al.,  2008, GCN
  Circ., 7777

\bibitem[\protect\citeauthoryear{{Roming}, {Kennedy}, {Mason}, {Nousek}, {Ahr},
  {Bingham}, {Broos}, {Carter} \& {et~al.}}{{Roming}
  et~al.}{2005}]{roming05short}
{Roming} P.~W.~A. et~al., 2005,
  Space Science Reviews, 120, 95

\bibitem[\protect\citeauthoryear{{Smith}, {Bradt} \& {Levine}}{{Smith}
  et~al.}{1999}]{smith99}
{Smith} D.~A.,  {Bradt} H.~V.,    {Levine} A.~M.,  1999, \apjl, 519, L147

\bibitem[\protect\citeauthoryear{{Tiengo}, {Rea}, {Klein-Wolt}, {Wijnands},
  {Israel}, {Esposito}, {Zane}, {Starling}, {Mereghetti}, {De~Luca} \&
  {G\"otz}}{{Tiengo} et~al.}{2008}]{tiengo08atel1559}
{Tiengo} A. et~al.,  2008, Astron. Tel., 1559

\bibitem[\protect\citeauthoryear{{Thompson} \& {Duncan}}{{Thompson} \&
  {Duncan}}{1995}]{thompson95}
{Thompson} C.,  {Duncan} R.~C.,  1995, \mnras, 275, 255

\bibitem[\protect\citeauthoryear{{Thompson}, {Lyutikov} \&
  {Kulkarni}}{{Thompson} et~al.}{2002}]{tlk02}
{Thompson} C.,  {Lyutikov} M.,    {Kulkarni} S.~R.,  2002, \apj, 574, 332

\bibitem[\protect\citeauthoryear{{Wachter}, {Patel}, {Kouveliotou}, {Bouchet},
  {{\" O}zel}, {Tennant}, {Woods}, {Hurley}, {Becker} \& {Slane}}{{Wachter}
  et~al.}{2004}]{wachter04}
{Wachter} S. et~al., 2004, \apj, 615, 887

\bibitem[\protect\citeauthoryear{{Woods}, {Kaspi}, {Thompson}, {Gavriil},
  {Marshall}, {Chakrabarty}, {Flanagan}, {Heyl} \& {Hernquist}}{{Woods}
  et~al.}{2004}]{woods04}
{Woods} P.~M. et~al.,
  2004, \apj, 605, 378

\bibitem[\protect\citeauthoryear{{Woods}, {Kouveliotou}, {G{\"o}{\u g}{\"u}{\c
  s}} \& {Hurley}}{{Woods} et~al.}{2008}]{woods08atel1549}
{Woods} P.~M.,  {Kouveliotou} C.,  {G{\"o}{\u g}{\"u}{\c s}} E.,    {Hurley}
  K.,  2008, Astron. Tel., 1549

\bibitem[\protect\citeauthoryear{{Woods}, {Kouveliotou}, {G{\"o}{\u g}{\"u}{\c
  s}}, {Hurley} \& {Tomsick}}{{Woods} et~al.}{2008}]{woods08atel1564}
{Woods} P.~M.,  {Kouveliotou} C.,  {G{\"o}{\u g}{\"u}{\c s}} E.,  {Hurley} K.,
    {Tomsick} J.,  2008, Astron. Tel., 1564

\bibitem[\protect\citeauthoryear{{Woods}, {Kouveliotou}, {van Paradijs},
  {Hurley}, {Kippen}, {Finger}, {Briggs}, {Dieters} \& {Fishman}}{{Woods}
  et~al.}{1999}]{woods99}
{Woods} P.~M. et~al., 1999, \apjl, 519, L139

\end{thebibliography}
\bsp

\label{lastpage}

\end{document}